\documentclass[prd,twocolumn,aps,amsmath,amssymb,nofootinbib,
preprintnumbers,superscriptaddress]{revtex4}

\IfFileExists{srcltx.sty}{\usepackage[active]{srcltx}}

\usepackage{graphicx}
\usepackage{dcolumn}
\usepackage{bm}
\usepackage{amsmath}
\usepackage{amsfonts}
\begin{document}

\input epsf \renewcommand{\topfraction}{0.8}

\preprint{UCLA/08/TEP/25}

\title{Gravitational waves from fragmentation of a primordial
scalar condensate into Q-balls}
\author{Alexander Kusenko}
\affiliation{Department of Physics and Astronomy, University of California, Los
Angeles, CA 90095-1547, USA}
\author{Anupam Mazumdar}
\affiliation{Physics Department, Lancaster University, Lancaster, LA1 4YB, UK}
\affiliation{Niels Bohr Institute, Blegdamsvej-17, Copenhagen, DK-2100, Denmark
}


\begin{abstract}
A generic consequence of supersymmetry is formation of a scalar condensate   
along the flat directions of the potential at the end of cosmological
inflation. This condensate is usually unstable, and it can fragment into
non-topological solitons, Q-balls.  The gravitational waves produced by the
fragmentation can be detected by Laser Interferometer Space Antenna (LISA), 
Advanced Laser Interferometer Gravitational-Wave Observatory (LIGO), and Big Bang Observer (BBO), 
which can offer an important window on the early universe and the
physics at some very high energy scales. 

\end{abstract}


\maketitle


Supersymmetry is widely regarded as a likely candidate for physics beyond the
Standard Model.   While many variants of supersymmetry have been considered,
all of them have scalar potentials with some flat directions lifted only by the 
supersymmetry-breaking terms.  At the end of cosmological inflation, the
formation of a scalar condensate along the flat directions can have a number of
important consequences~\cite{reviews}.   In particular, it can be responsible
for generation of the matter-antimatter asymmetry via 
Affleck--Dine (AD) mechanism~\cite{AD}, and, in some models, dark matter can be 
produced in the same process~\cite{KS,dark_matter}.  Some flat directions could
be responsible for the primordial inflation~\cite{Randall:1995dj,AEGM}. 

The formation of AD condensate is a generic phenomenon, relying only
on the assumptions of inflation and supersymmetry.   In general, this condensate
is unstable: an initially homogeneous condensate can break up into lumps of
the scalar field, called Q-balls~\cite{Coleman}, under some very generic 
conditions~\cite{KS}.  All phenomenologically acceptable supersymmetric
generalizations of the Standard Model admit Q-balls ~\cite{Kusenko:1997zq},
which can be stable, or can decay into
fermions~\cite{Kusenko:1997zq,DKS}. The formation of Q-balls is accompanied by a
coherent motion of the scalar condensate, which creates the source of gravity
waves.  We will show that fragmentation of the scalar condensate into Q-balls
can produce gravitational waves detectable by LISA~\cite{LISA}, 
LIGO~III~\cite{LIGOIII}, and BBO~\cite{BBO}.

The physics of AD condensate fragmentation has been studied both
analytically~\cite{KS,EMc,EJM,Pawl,Johnson:2008se} and
numerically~\cite{Minos,KK,MV,Volkov,Palti,Broadhead,Campanelli}.  At
the end of inflation (assuming that the inflation occurs in a hidden sector at high scales)
the condensate has a uniform density, with small
perturbations of the order of $10^{-5}$~\cite{reviews}.   Under some rather
generic conditions, the instabilities develop and lead, eventually, to the
formation of Q-balls, which can either decay or remain as stable
relics~\cite{KS,EMc}.  Although the final
product of such evolution, Q-balls in the ground state,  are spherically
symmetric, the coherent motions associated with the condensate
fragmentation and re-arrangement are not spherically symmetric.  Moreover, the
newly formed Q-balls first appear in their excited states and oscillate until
they settle in the spherically symmetric ground states~\cite{KK,MV}.   The lack of
spherical symmetry in the process of fragmentation is essential for generating
the gravity waves.  

Following the general picture developed in Refs.~\cite{KS,EJM}, the 
scalar condensate undergoing fragmentation can be approximated, in the linear
regime, as $\phi(x,t)=\phi(t) \equiv R(t) e^{i \Omega(t)}$, plus a 
perturbation $\delta R, \delta \Omega \propto e^{S(t) - i \vec{k}\vec{x}} $.
One finds that the homogeneous solution is unstable due to some exponentially 
growing modes, ${\rm Re} \, \alpha>0$, where $\alpha = dS/dt
$~\cite{KS,EJM,Johnson:2008se}. The mass density of the condensate undergoing
fragmentation can be written as 
\begin{equation}
 \rho(x,t) = \rho_0 + \rho_1(x,t)\,, 
\end{equation}
where
\begin{equation}
\rho_1(x,t)=  \epsilon \rho_0 \int d^3k  \,e^{\alpha_k t} \cos(\omega t-\vec{k}\cdot
\vec{x}) \,.
\end{equation}
The instability develops when there is a band  of growing modes with positive
and large enough $\alpha_k$~\cite{KS}.  The linear approximation breaks down
when $\epsilon \exp ( \alpha_k t) \sim 1$, but we will use this representation, 
up to its limit of applicability, to get the estimates of the gravity waves
produced.  

The quadrupole moment that generates gravity waves is given by~\cite{Weinberg}
\begin{equation}
 D_{ij} = \int d^3x\ x_i x_j \,  T^{00}(x,t)\,,
\end{equation}
where the energy-momentum tensor $T^{00}(x,t)\approx \rho(x,t)$. The space
integration is over some arbitrary volume.  

The power emitted in gravity waves in one frequency mode is given by:
\begin{equation}
P(\omega)=\frac{2}{5} G \omega^6 \left( D^*_{ij}(\omega)
D_{ij}(\omega)-\frac{1}{3}|D_{ij}(\omega)|^2\right)\,,  
\end{equation}
and the total energy emitted in gravitational waves, in all
frequencies, is given by:
\begin{equation}
E \sim  \left( 2\pi \int p(\omega) d\omega \right) \times \Delta t\,,
\end{equation}
where $\Delta t$ is the duration of the fragmentation. 

Based on the analytical and numerical calculations of the condensate
fragmentation~\cite{KS,KK,Kasuya2}, we take the typical parameters of the
fastest-growing mode:
\begin{equation}
 k \sim \xi_k\times 10^2 H_{\ast}, \  \omega_k \sim v k \sim \xi_k\times 10^2 \, v H_{\ast}, 
\end{equation}
where $H_{\ast}$ is the Hubble constant at the time of the condensate
fragmentation, and $v$ is the typical group velocity of the wave front in the
evolution of the condensate, and we expect that the dimensionless factor 
$\xi_k \sim 1$, based on  the results of Refs.~\cite{KS,KK,Kasuya2}. 

Since no cancellations are expected in the absence of spherical symmetry,
we replace the $x_ix_j$ by $( f_k\times 10^2 H_{\ast})^{-2}$ in the space integration, 
take the volume to be $V\sim H^{-3}_{\ast}$, and assume that $\epsilon
\exp(\alpha_k t) \sim 1$. Then, for the leading mode, 
\begin{equation}
 D_{ij}(t) \sim H_\ast^{-3} \left( 10^2 H_{\ast} \right)^{-2} \rho_0
\cos(\omega_k 
t-kx)\,, 
\end{equation}
and, in frequency space, 
\begin{equation}
 D_{ij}(\omega) \sim 10^{-4}  \xi_k^{-2}\frac{\rho_0}{H_{\ast}^5} \,.
\end{equation}
For $\omega \sim 10^2 v H_{\ast}$, we estimate  the power in gravitational
waves in a Hubble volume: 
\begin{equation}
P\sim 10^{4} \xi_k^{-2} \, G \frac{\rho_0^2 v^6}{H^4_{\ast}}\,.
\end{equation}
To estimate the velocity of the wavefront in the process of fragmentation, we
note that, for the mode $\phi(x,t) \approx R(t) \exp\{ \alpha_k t\} \cos
(\omega_k t - kx)$, where $R(t)$ is a slowly changing function of
time,\footnote{The adiabatic limit $\dot{R}/R \rightarrow 0$ is amenable to
perturbation theory~\cite{KS}. This case corresponds to a large global charge
density.  For a rapidly varying $R(t)$, the numerical calculations give similar
results regarding the fragmentation time and length scales~\cite{KK}.}
\begin{equation}
 v\sim \xi_v \left | \dot{\phi}/\phi'_x \right |\sim \xi_v\ \alpha_k/k, \ \ \xi_v \sim 1,
\end{equation}
where the uncertainty factor $\xi_v \sim 1$ will be retained to keep track of the uncertainty in the final answer.  

The relation between $\alpha_k$ and $k$ is given by a dispersion 
relation~\cite{KS}, which takes a simple form, 
\begin{equation}
(\alpha_k^2+k^2) \left (\alpha_k^2- \left (\dot{\Omega}^2-V''(R) \right) \right )
+4\dot{\Omega}^2 \alpha_k^2=0,
\end{equation}
under the following assumptions: $H\ll k\sim \alpha_k\ll
\sqrt{\dot{\Omega}^2-V''(R)}\sim \dot{\Omega}
\sim m_\phi$ valid in the case of the fastest-growing mode in gravity mediated
supersymmetry breaking models (the latter is essential for the $\dot{\Omega}\sim
m_\phi$ condition~\cite{reviews}).  This equation has an approximate solution: $
\alpha_k \approx k/\sqrt{3} $, and 
\begin{equation}
 v^6 \sim\xi_v^6 (\alpha_k/k)^6 \sim \xi_v^6 (1/\sqrt{3})^{6} \sim 10^{-2} \xi_v^6 \,.
\end{equation}
The fragmentation takes place on the time scale of
the order of $\Delta t \sim \alpha_{k}^{-1}\sim \xi_k^{-1} 10^{-2}H^{-1}_{\ast}$.  (Here we
neglect the possible contributions from the collisions and oscillations of
Q-balls, which can take place on a much longer time scale~\cite{KS,MV1}.) The
total energy in gravity waves generated in the Hubble volume is
\begin{equation}
E\sim P \Delta t \sim  G \frac{\rho_0^2}{H_{\ast}^5}\, \xi_k^{-3} \xi_v^6\,.
\label{E_produced}
\end{equation}
This corresponds to the energy density in gravitational waves at the time of
production 
\begin{equation}
\rho_{GW\ast} \sim  10^{-3} \xi_k^{-3} \xi_v^6\,
\frac{\rho_0^2 }{H^2_{\ast} M_{\rm Pl}^2}\,,
\label{rho_produced}
\end{equation}
where $M_{\rm Pl}=1/\sqrt{8\pi G}$ is the reduced Planck mass. 
Hence,  the fraction of the energy density in gravitational waves at the
time of production is 
\begin{equation}
\Omega_{GW*}  \sim 10^{-3} \xi_k^{-3} \xi_v^6\,\frac{\rho_0^2}{(H_{\ast} 
M_{\rm Pl})^4}\,.
\end{equation}
If the energy density of the condensate is comparable to the total energy
density, or if the condensate energy dominates the energy in the universe,
then $\rho_0\sim 3H_{\ast}^2M_{\rm Pl}^2$, and $\Omega_{GW*}\sim 10^{-3}$.  

The energy density in the condensate depends on the model, and, foremost, 
on the type of supersymmetry breaking terms that lift the flat direction.
This is because the potential along the flat direction depends on supersymmetry
breaking (it vanishes in the limit of exact supersymmetry), and there are many
ways to break supersymmetry. In gauge-mediated supersymmetry breaking scenarios
the potential can have the form~\cite{KS}
\begin{equation}
V(\phi)\approx M_{S}^4\log\left(1+\frac{|\phi|^2}{M_{S}^2}\right)\,.
\label{gaugemediatedsusybreaking}
\end{equation}
Here $M_S$ is the scale of supersymmetry breaking, which is of the order of
${\cal O}(1)$~TeV.  In gravity mediated scenarios, the flat
directions are lifted by mass terms that persist all the way to the
Planck scale~\cite{EMc}:
\begin{equation}
V(\varphi) \approx m_{\phi}^2 \left(1+K\log\left(\frac{|\phi|^2}{M_{Pl}^2}
\right)\right) |\phi|^2\,,
\label{gravitationalsusybreaking} 
\end{equation}
where $K\sim 0.05$ (for squark directions) describes the running of
the mass term~\cite{EMc}. Since $m_\phi \sim M_S\sim 1-10$~TeV in typical
models, both potentials are phenomenologically acceptable near the minimum. 
However, in AD condensate and inside the Q-balls that form in its fragmentation,
the vacuum expectation value (VEV) can be very large.  

The main difference between gauge and gravity mediated cases for us is the mass
per baryon number stored in the AD condensate and in the Q-balls that form
eventually as a result of the fragmentation. In the gravity mediated scenarios,
the mass density is $ \rho_0 \sim m_\phi^2 \phi^2 $, the global charge
density is $ n_Q\sim m_\phi \phi^2$, and the mass per unit global charge is of
the order of $m_\phi$, independent of the VEV $\phi_0$. In gauge mediated
scenarios, the mass density is $ \rho_0 \sim m_\phi^4$,  the global charge
density is $ n_Q\sim m_\phi \phi^2$, and the mass per unit global charge is
$\rho_0/n_Q\sim m_\phi^2/\phi$~\cite{KS,DKS}. 

The flat directions that carry a non-zero global charge $Q=(B-L)$ contribute to
the generation of the baryon asymmetry via AD process~\cite{AD}.  The
requirement that $\eta_B=n_B/n_\gamma \sim 10^{-10}$ implies that the total mass
density of such a condensate cannot be of the order of the total density of the
universe in generic models.  This is true in both gauge-mediated and
gravity-mediated supersymmetry breaking models.  The gravity waves from the
fragmentation of such a condensate are well below the capabilities of the
current and planned detectors.

On the other hand, there are flat directions whose  baryon number $B$ and
lepton number $L$ are equal to each other~\cite{DRT,GKM,reviews}.  While the
flat directions with $B\neq L$ contribute to baryon asymmetry of the universe,
those with $B=L$, have zero $(B-L)$ density.   Electroweak sphalerons destroy
any primordial $(B+L)$ asymmetry, and so the corresponding $\eta_{B+L}=
n_{B+L}/n_\gamma$ is not constrained.  It is possible that, at the time of
the  fragmentation,  $\eta_{B+L} \gg \eta_B $.  For $B=L$ flat directions,
there is no reason why $\rho_0$ cannot be of the order of the total energy
density.  The fragmentation of such flat directions can produce a detectable
level of gravitational waves. 

There are various examples in the literature of the flat directions that can
dominate the energy density of the universe, while they do not contribute to
(and are not constrained by) the observed baryon asymmetry of the universe. 
This is the case, for example, when the effective mass for the phase direction
is large during inflation, which results in the initial condition with a very
small $\dot \Omega $ for the scalar condensate $\phi(x,t)=\phi(t) \equiv R(t)
e^{i \Omega(t)}$.  This is also the case  when  the inflation is driven by a
flat direction,  $udd$, $LLe$ or $NH_uL$~\cite{Randall:1995dj,AEGM,AEGJM}.
Another well-know example is a flat direction that acts as a curvaton and
dominates the energy density of the universe at the time of oscillations and
decay~\cite{Curvaton}. In all of these cases, the net global charge of the
condensate is negligible, but the fragmentation can still occur and produce
Q-balls and anti-Q-balls~\cite{Kasuya1,Kasuya2}.

Once the gravitational waves are created, they are decoupled from the rest of
the plasma. We can estimate the peak frequency of the gravitational
radiation observed today, $f_{\ast}=\omega_{k}/2\pi$:
\begin{eqnarray}
\label{peak}
f&=& f_{\ast}\frac{a_{\ast}}{a_0}=f_{\ast}\left(\frac{a_{\ast}}{a_{\rm rh}}\right)
\left(\frac{g_{s,0}}{g_{s,{\rm rh}}}\right)^{1/3}
\left(\frac{T_{0}}{T_{\rm rh}} \right)\, \\
&\approx & 0.6 \, {\rm
mHz}\ \xi_k\xi_v\, \left(\frac{g_{s,{\rm rh}}}{100}\right)^{1/6}\left(\frac{T_{\rm rh}}
{1~{\rm TeV}}\right) \left(\frac{f_{\ast}}{10 H_{\ast}}\right)\,, \nonumber 
\end{eqnarray}
where we have assumed that $a_{\ast}\approx
a_{\rm rh}$, which also means that 
during the oscillations of the AD condensate the effect of the Hubble expansion
is negligible. The values of relativistic degrees of freedom are $g_{s,{\rm
rh}}\sim 300$, $g_{s,0}\sim 3.36$.  The subscript ``rh'' denotes the
epoch of reheating and thermalization, while the subscript ``0'' refers to the
present time.  As we discussed, the typical frequency of the oscillations of the
AD condensate is  $\omega_{k}\sim 10^2 H_{\ast}$~\cite{KS,KK,Kasuya2}. Then for
$T_{\rm rh}\sim 1$~TeV  (such a value of the reheat temperature is natural when 
the flat direction is responsible for reheating the universe~\cite{Rouzbeh,AEGJM}), 
the frequency is of the order of mHz,  which is in the right frequency range
for LISA~\cite{LISA}.  A higher temperature $T_{\rm rh}\sim100$~TeV corresponds 
to the  LIGO III frequency range, $10-100$~Hz~\cite{LIGOIII}.  Signals in both of these ranges will be accessible to BBO~\cite{BBO}.  Since the supersymmetry breaking scale is related to the energy in the condensate, as well as the reheating temperature,  LIGOIII and BBO could be in the position to probe supersymmetry broken above 100~TeV, beyond the reach of Large Hadron Collider (LHC).  

The fraction of the critical energy density $\rho_c$ stored in the gravity waves
today is
\begin{eqnarray}
\label{maxamp}
\Omega_{GW} & = &\Omega_{GW^{\ast}}
\left(\frac{a_{\ast}}{a_0}\right)^4\left (\frac{H_{\ast}}{H_0}\right)^2 \\ 
&\approx&  \frac{1.67\times 10^{-5}}{ h^{2}}\left(\frac{100}{g_{s,\ast}}\right)^{1/3}
\Omega_{GW^{\ast}}  
\approx 10^{-8} \, \xi_k^{-3} \xi_v^6\, h^{-2} \nonumber 
\end{eqnarray}
where $a_0$ and $H_0$ are the present values of the scale factor and the Hubble 
expansion rate.  LISA can detect the gravitational waves down
to $\Omega_{GW}h^2\sim 10^{-11}$ at mHz frequencies, while LIGO~III is sensitive
to  $\Omega_{GW}h^2 \sim(10^{-5}-10^{-11})$ 
in the $(5-10^{3})$~Hz frequency band.  Therefore,  the gravitational 
waves with $\Omega_{GW*}\sim 10^{-3}$ and $\Omega_{GW}h^2\sim 10^{-8}$ from the 
fragmentation of an AD condensate can be detected.   The first results of our numerical simulations 
(work in progress) appear to produce the gravitational wave signal that is somewhat weaker.  We attribute the difference to the value of the uncertainty factor  $\xi_k^{-3} \xi_v^6$, which is especially sensitive to the average wavefront velocity.

As one can see from eq.~(\ref{rho_produced}), the power generated in frequency $\omega$ is proportional to $$\xi_k^{-3}\sim \left( \frac{\omega}{10^{2} H_*}\right)^{-3}. $$  Hence, the spectrum is strongly peaked near the longest wavelength, of the order of the Q-ball size~\cite{KS,KK,Kasuya2}.  The relatively narrow spectral width will help distinguish this signal from the gravity waves generated by inflation~\cite{inflation}, which are expected to have an approximately scale-invariant spectrum (and a smaller amplitude).  Future numerical simulations will help refine the prediction for the signal from Q-ball formation, which can help distinguish this source from a phase transition in the early universe~\cite{Kamionkowski:1993fg}.  LISA and LIGO~III will be able to discriminate the gravity waves due to fragmentation from those of point sources, such as merging black holes and neutron stars, which have  specific ``chirp'' properties~\cite{Owen:1998dk}.  Furthermore, the signal discussed here will not create a significant background for the cosmic microwave polarization experiments, such as B-Pol~\cite{B-pol}, which can detect the gravity waves with extremely long wavelength.

Some additional gravitational waves can be generated by
collisions and oscillations of Q-balls~\cite{KS,MV1}.  
We leave the discussion of the magnitude of this additional contribution to
future studies. 

To summarize, the fragmentation of a scalar condensate into Q-balls, which is a
generic consequence of supersymmetry and inflation, can produce a detectable
level of gravitational waves, up to  $\Omega_{GW}h^2\sim 10^{-8}$, near the peak
frequency of BOO and either LISA or LIGO, depending on the reheating
temperature.  Detection of the gravitational waves form this process can shed
light on the earliest post-inflationary epoch in the history of the universe, can probe supersymmetry even if it is broken at a scale above 100 TeV, and can provide information about new physics at some very high energy scales associated with the flat directions.  

A.K. thanks S.~Phinney for helpful discussions.  The work of A.K.  was supported
in part by the DOE grant DE-FG03-91ER40662 and by the NASA ATFP grant 
NNX08AL48G.  Research of A.M. is supported in part by the grant
MRTN-CT-2006-035863.  A.K. thanks Aspen Center for Physics for hospitality.



\end{document}